\documentclass[namedreferences]{SolarPhysics}
\usepackage[optionalrh]{spr-sola-addons} 
\usepackage{graphicx}        
\usepackage{color}           
\usepackage{url}             

\newcommand{\aap}{    {\it Astron. Astrophys.}}

\newcommand{\aj}{     {\it Astron. J.}}
\newcommand{\apj}{    {\it Astrophys. J.}}

\newcommand{\grl}{    {\it Geophys. Res. Lett.}}

\newcommand{\mnras}{  {\it Mon. Not. Roy. Astron. Soc.}}

\newcommand{\solphys}{{\it Solar Phys.}}

\newcommand{\memsai}  {\it Mem. Soc. Astron. Ital.}
\newcommand{\caii}     {Ca {\sc ii}~K~}
\newcommand{\caiih}     {Ca {\sc ii}~H~}
\newcommand{\fei}     {Fe {\sc i}~}
\newcommand{\tiii}     {Ti {\sc ii}~}
\newcommand{\ci}     {C {\sc i}~}
\newcommand{\mni}     {Mn {\sc i}~}
\begin{document}
\begin{article}
\begin{opening}
\title{Improvements in the determination of ISS \caii parameters\\ {\it Solar Physics}}
\author{L.~\surname{Bertello}$^{1}$\sep
       A.A.~\surname{Pevtsov}$^{2}$\sep
       J.W.~\surname{Harvey}$^{1}$\sep
       R.M.~\surname{Toussaint}$^{1}$
       }
\runningauthor{Bertello {\it et al.}}
\runningtitle{ISS Ca II K parameters}
\institute{
            $^{1}$ National Solar Observatory, 950 North Cherry Avenue, Tucson, AZ, USA
            $^{2}$ National Solar Observatory, Sunspot, NM, 88349, USA \\
   email: \url{bertello@noao.edu} email: \url{pevtsov@noao.edu}
   email: \url{jharvey@noao.edu}  email: \url{roberta@noao.edu} \\ 
             }
\begin{abstract}
Measurements of the ionized \caii line are one of the major resources
for long-term studies of solar and stellar activity. They also play a critical
role in many studies related to solar irradiance variability, particularly
as a ground-based proxy to model the solar ultraviolet flux variation that
may influence the Earth's climate. 
Full disk images of the Sun in \caii have been available from various observatories for more than
100 years and latter synoptic Sun-as-a-star observations in \caii began in the early 1970s. 
One of these instruments, the Integrated Sunlight Spectrometer (ISS) has been in operation
at Kitt Peak (Arizona)
since late 2006. The ISS takes daily observations of solar spectra in nine spectra bands, including the \caii and H lines. 
We describe recent improvements in data reduction of \caii observations, and present time
variations of nine parameters derived from the profile of this spectral line. 
\end{abstract}
\keywords{Solar Activity, Observations, Data Analysis; Chromosphere, Active}
\end{opening}
\section{Introduction}

The diagnostic value of \caii measurements for the study of solar
magnetism and irradiance variability is well established (e.g., \opencite{2009SoPh..255..229F} and references therein). These measurements provide information on energy input to solar chromosphere and its long-term variation. 
The intensity of the K-line is a proxy for total magnetic flux (e.g., \opencite{2005MmSAI..76.1018O}), 
and it can be used to represent solar variability in (extreme)  ultraviolet, (E)UV 
(e.g., \opencite{2002GeoRL..29w...4F}). Solar observations are also used to interpret and calibrate 
stellar measurements of Ca II K and H spectra in terms of stellar activity and cycles (e.g.,
\opencite{1984A&A...130..353R}; \opencite{2006ApJ...651..444G}; \opencite{2009MNRAS.398.1495V};
\opencite{2011AJ....141...34B}; \opencite{2010MNRAS.408.2290G}). As shown in some studies
(e.g. \opencite{1995SoPh..159...53D}; \opencite{2002AN....323..123H}), the connection
between the modulation of the disk-integrated \caii-line flux and solar surface differential rotation
may open the possibility of using these measurements to determine the differential rotation of other
main-sequence stars.

In a broader context, \caii observations have been used
to estimate distances to early-type stars (\opencite{2009A&A...507..833M}), and
as diagnostics of velocity fields in red giant stars (\opencite{2011A&A...526A...4V}).
It has also been suggested to use recurring chromospheric emission visible in \caii as a tool
to detect massive extra-solar planets orbiting very close to their central stars 
(\opencite{2010arXiv1012.1720L}). 

Since 1974, a long-term program of monitoring the sun-integrated \caii line has been operating at the National 
Solar Observatory at Kitt Peak (\opencite{1977ApJ...211..281L}; 
\opencite{2007ApJ...657.1137L}) and since 1977 at Sacramento Peak 
(\opencite{1984ApJ...276..766K}; \opencite{1998ASPC..140..301K}). 
Starting December 2006, disk-integrated sunlight spectra in \caii line 
are recorded on a daily basis by the Integrated Sunlight Spectrometer (ISS). The ISS
is one of three instruments, 
with the Vector Spectro-Magnetograph (VSM) and the Full Disk Patrol (FDP), 
comprising the Synoptic Optical Long-term Investigations of the Sun (SOLIS) - a synoptic facility for solar 
observations operating at NSO/Kitt Peak (\opencite{2003SPIE.4853..194K}; 
\opencite{2003ASPC..307...13K}). 
A Solar-Stellar Spectrograph project started at Lowell
Observatory also provides regular observation of the solar K line since 1994
(\opencite{1998ApJ...493..494H}; \opencite{2007AJ....133..862H}).

Currently the ISS instrument provides measurements in nine different spectral bands, including the \caii band 
centered at 393.37 nm. \inlinecite{2010MmSAI..81..643L} have shown that the ISS Ca K index compares well with
both the \caii McMath-Pierce spectral line scans and the 10.7 cm radio flux.
Here we provide the first detailed description of the data reduction and describe recent improvements 
in the determination of the nine different 
parameter time series derived from the \caii spectra.
These improvements result from changing most of the original ISS code that processes the observed 
\caii line profiles and then computes the parameters.
The improvement is more
significant for some of the parameters, while others benefit only marginally from
these changes.
Sections 2 and 3 provide a brief description of the ISS instrument, the observing sequence, and the data reduction. 
In Section 4, we describe the derivation of nine parameters representing properties of \caii line profiles, and 
compare the results of improved data analysis with previous data reduction. 
Section 5 provides a brief summary of the results of this study.

\section{The ISS Instrument}

The Integrated Sunlight Spectrometer(ISS) operating at the National Solar Observatory at Kitt Peak (Arizona)
is designed to obtain high spectral resolution (R $\cong$ 300,000) observations of the Sun as a star in 
a broad range of wavelengths of 350 nm -1100 nm. 
Beginning December 1, 2006, the ISS takes daily observations in nine spectral bands centered at 
CN head band 388.40 nm, \caiih (396.85 nm), \caii (393.37 nm), \ci 538.00, \mni 539.41 nm, 
H-alpha 656.30 nm, Ca {\sc ii}~ 854.19 nm, He {\sc i}~ 1083.02 nm, and NaD1 (589.59 nm). 
In addition, high resolution spectra formed by the sunlight passing through an iodine vapor cell were 
recorded beginning January 7, 2008. Iodine spectra are over imposed on the solar spectral range 
centered at \ci 538.00 nm. Table \ref{iss} lists currently observed spectral lines.

\begin{table}
\caption{Spectral bands measured by the ISS instrument. Each spectral band is centered
at $\lambda_0$, with a bandwidth given by $\Delta\lambda$. The average linear dispersion
over time, $d\lambda/dx$, is also given. The start date (fifth column) indicates
the first day ISS began to observe in that particular spectral band.}
\begin{tabular}{lllll}
\hline
Type & $\lambda_0$, nm & $\Delta\lambda$, nm & $d\lambda/dx$, pm/pixel & Start date \\
\hline
CN band & 388.40 & 0.58 & 0.564 & December 4, 2006 \\
\caii & 393.37 & 0.55 & 0.541 & December 1, 2006 \\
\caiih & 396.85 & 0.53 & 0.522 & December 4, 2006 \\
\ci & 538.00 & 0.84 & 0.824 & December 4, 2006 \\
\ci (with iodine lines) & 538.00 & 0.84 & 0.824 & January 7, 2008 \\
\mni & 539.41 & 0.83 & 0.816 & December 4, 2006 \\
H-alpha & 656.30 & 1.14 & 1.12 & August 31, 2007 \\
Ca {\sc ii} & 854.19 & 1.61 & 1.58 & December 13, 2006 \\
He {\sc i} & 1083.02 & 1.65 & 1.61 & December 4, 2006 \\
NaD1 & 589.59 & 0.98 & 0.956 & March 23, 2011 \\
\hline
\end{tabular}

\label{iss}
\end{table}

Observations of integrated sunlight with the ISS are accomplished through  the use of a fiber optic feed. 
A small optical system (a lens of 8-mm diameter) installed on the side of main mount of SOLIS/VSM focuses a 
400 micron diameter image of the Sun on the input face of a 600 $\mu$ diameter fiber. 
The fiber assembly transmits light 
to a McPherson 2-m Czerny-Turner double-pass spectrograph located in a temperature-controlled room below the telescope. 
The output beam from the fiber consists of sunlight scrambled in both angle and position so that any output 
angle from any position on the exit face is well-integrated. A prism pre-disperser isolates the desired 
wavelength band. The spectrograph employs a 316 g/mm grating blazed at 63.5 degrees and a movable, 
back-illuminated 512 x 1024 CCD as the focal plane detector. 
Additional information, including transmission characteristics of the fiber optics, the instrument's block diagram, 
mechanical and optical layout are available from the SOLIS web site at http://solis.nso.edu/ISSOverview.html. 
Daily observations can be accessed from http://solis.nso.edu/iss.

\section{Data Processing}

For each spectral band, observations are taken in four different positions by moving the CCD
in respect to the spectra in the spectrograph focal plane: the first image (image
1, 512 x 1024 pixels in size) is taken with the CCD centered at a corresponding
wavelength (for a given spectral band), image 2 is shifted by 1 pixel in the
direction of dispersion, image 3 is shifted by 11 pixels, and image 4 by 129
pixels. It takes about 5 minutes to complete the full cycle of observations for
most spectral bands; for \caii observations it takes about 12 minutes, and for
He {\sc i}~ 1083.0 nm about 15 minutes. Four images are used both for the purpose
of flat fielding and to create the final spectrum for a given spectral band. Because
in four images the CCD is shifted in respect to fixed spectral features, the same pixels
are exposed to a different level of light. Assuming that the solar source remains 
constant during the observing cycle (3 - 15 minutes), the images can be used to derive flatfield and CCD gain based
on an algorithm described in \inlinecite{2003AJ....126.1112T}. A dark field
is taken for each observed spectral range at the beginning of four-image cycle by closing 
the intermediate shutter of the double pass system.
The same dark field is used for all four images taken in each spectral band.

After applying dark and flat fields, spectral features on four spectral images
are aligned, and the astigmatic images are averaged in the spatial direction to create
the final spectral line profile for a given spectral band. In this averaging, each
spatial spectrum of the four 512 x 1024 pixel images is weighted by the square root of
its total intensity. The final products are intensity as a function of
pixel number (spectral line profile) and its error for each pixel (computed as
standard deviation). At the next step of data reduction, the pixel position of selected
spectral lines is determined. Pixel positions are converted to nanometers using
known wavelengths of selected spectral lines from high resolution spectra taken
with the NSO Fourier Transform Spectrometer (FTS,  \opencite{2007assp.book.....W}).

Spectra reduced to this level still show a slight gradient in intensity in
wavelength direction. For spectral bands that include continuum, this gradient is
removed by a linear fit to the continuum and scaling the spectra
by the fit. For the \caii and H lines, ISS observations do not include
continuum. These two spectral bands are scaled using intensities at two narrow
spectral bands situated in blue and red wings of these lines.  Mean intensities
in these two bands are scaled to match spectral line profiles taken by the FTS.
For \caii spectra these two bands are: 393.147-393.153 nm in the blue wing, and
393.480-393.500 nm in the red wing. The intensity profile is then scaled so that
the mean residual intensity between 393.4958-393.5134 nm is 0.1756. 

Reduced spectra are saved in FITS format as double array with three axes:
wavelength in \AA ngstroms, continuum normalized intensity, and error in intensity
for each pixel. These level 1 FITS files constitute the basic ISS data product.

\section{The ISS \caii Parameter Time Series}

The importance of \caii parameter time series for monitoring solar and stellar activity
has been established by many decades of studies (e.g. \opencite{1957ApJ...125..661W};
\opencite{1968ApJ...153..221W}; \opencite{1978ApJ...226..679W}).
Currently, nine parameters are routinely derived from each measured
393.37 nm ISS spectrum. The quality of these parameter-time series depends strongly
on our ability to measure the linear dispersion of the instrument over time.

The ISS Ca K spectral band covers slightly more than 0.5 nm, centered on K3.
Within this particular spectral range the ISS measures the profile of four neutral iron
lines and a singly ionized titanium line. Table \ref{tabdisp} lists these five spectral
lines and their corresponding wavelength, taken from \inlinecite{1974kptp.book.....P}. These
lines are used to determine the linear dispersion for the \caii range, as described below.

\begin{table}
\caption{Observed line profiles within the ISS Ca K band and corresponding wavelengths ($\lambda$)
used in the determination of the linear dispersion.}
\begin{tabular}{llllll}
\hline
Element & $\lambda$, nm & Element & $\lambda$, nm  & Element & $\lambda$, nm \\ 
\hline
\fei  & 393.11251 & \fei  & 393.26354 & \fei  & 393.58213 \\
\tiii & 393.20172 & \fei  & 393.53133 & \\
\hline
\end{tabular}
\label{tabdisp}
\end{table}

\subsection{Linear dispersion}

Each ISS measurement includes 1024 spectral samples covering the CaK spectral
band. Although the ISS is an extremely stable instrument, mechanical hysteresis in 
the grating positioning mechanism
may result in (sometimes significantly) changes in pixel position of spectra features. 
In order to address this issue, we first
find the location of the K3 line core and then we use the relative distance of the
other lines with respect to K3 to determine their positions. We use a quadratic
fit and seven points around the line minima to determine the line core position
of the five lines used for the calculation of the dispersion.

Although the
level 1 fits files provide the error in intensity for each pixel, we use
unweighted polynomial fits throughout the data reduction process. The main reason for
this choice is due to the large pixel-to-pixel variability of these errors, in some cases up to a factor
20 or more. When the uncertainty in intensity is 
included into the fit the fluctuations in the computed coefficients are significantly
larger than in the case when the fit is unweighted, contrary to the very high
stability of the ISS. It also produces
small pivot elements in the variance-covariance matrix associated with the observed points used
in the fit of some spectra that results in a poor determinations of the parameters. 

We compute the formal 1-$\sigma$ estimates of the returned parameters
from the unweighted fit under the assumption that our model for the data is correct. This
is certainly true for the high-quality data we are using. After the five line
positions (in pixel units) are determined, a simple linear fit with the nominal wavelengths listed in
Table \ref{tabdisp} is used to calculate the linear dispersion.
 
\begin{figure}
\begin{center}
\includegraphics[width=1.0\textwidth]{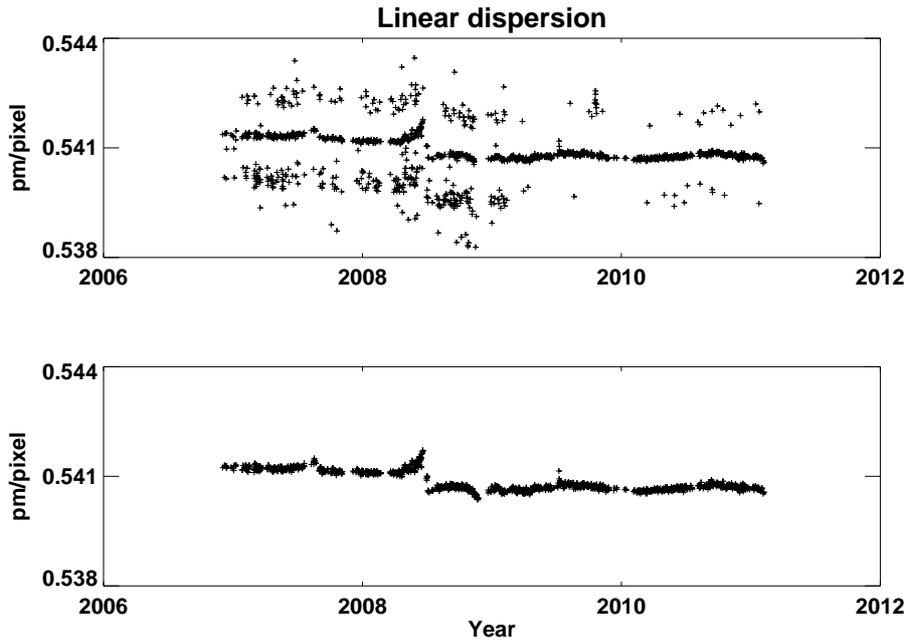}
\caption{Improved calculation of the linear dispersion for \caii observations
taken during the period December, 2006 - February, 2011. The new determination
of the five line core positions used to compute the dispersion clearly shows the
stability of the algorithm over time (bottom plot). The error bars are not shown for clarity (see text).}
\label{disp}
\end{center}
\end{figure}

Figure \ref{disp} shows the comparison between the linear dispersion calculated using
the new approach and the previous result from the ISS pipeline. A total of 1678 observations
has been used, that cover the period from December 2006 to February 2011. Unlike the old
reduction, all observations have now a well determined dispersion. The dispersion shows slight
($\sim$0.02\%) annual variations that may be attributed to change in the index of refraction of air.
A jump in the value of the dispersion occurring around June 2008 has been traced to a failure of a
thermostat that controls the temperature in the spectrograph room. 
After the thermostat was replaced, the spectrograph had stabilized at a slightly different dispersion. 
However, one should notice that this change in dispersion is extremely small, about 0.1\%. 

For purposes of clarity, we intentionally omitted the error bars in Figure \ref{disp} and all other figures
showing the temporal behavior of the parameters computed from the \caii profiles.
The errors on these parameters differ significantly depending on the strategy used to computed
the parameters. With the approach described in this paper the errors are generally
more uniform over time than the ones computed by the old ISS pipeline, in some cases smaller, and the 
corresponding error bars are typically
shorter than the size of the points shown in the plots.
We will discuss the errors in a separate section of this article.  

As part of this analysis we have investigated the temporal behavior of the core position
and intensity for the five spectral lines used to determine the linear dispersion. The core
positions of each line shows no significant temporal trends but the fluctuations around the
mean value, although very small, are quite different. These relative fluctuations vary from
$\sim 8 \times 10^{-9}$ (\tiii 393.20172) to $\sim 2 \times 10^{-7}$ (\fei 393.58213).
The core intensities of these spectral lines, 
on the other hand, show a tendency to increase over the time covered by this
investigation. This increase is less than 0.005 in residual intensity, for all lines.

\subsection{K3 wavelength position and intensity}

\begin{figure}
\begin{center}
\includegraphics[width=1.0\textwidth]{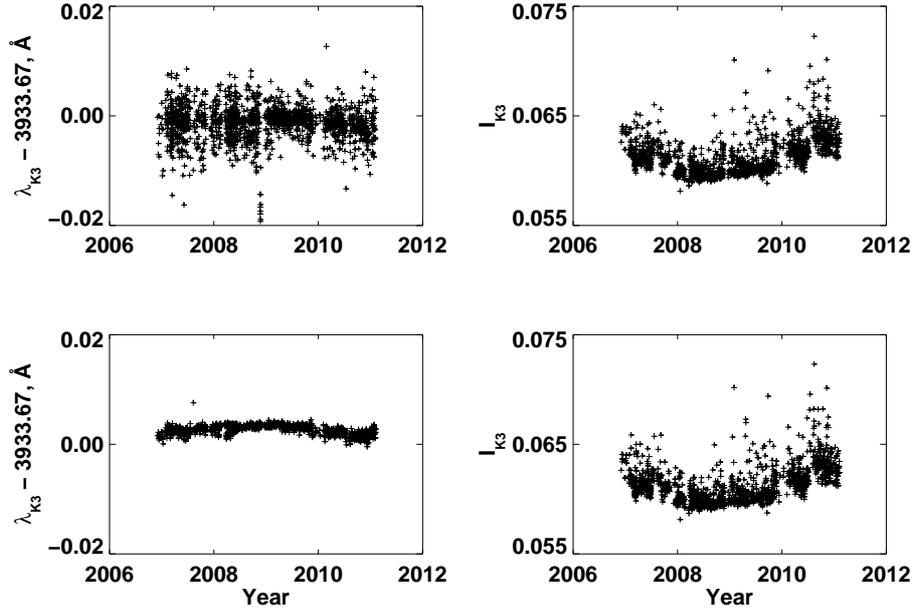}
\caption{Comparison between the $\lambda_{K3}$ and $I_{K3}$ time series computed from the
old ISS pipeline (top) and the new and improved version (bottom).
With the exception of very few data points, the impact of the new code in the computation of I$_{\rm K3}$
is negligible.
The error bars are not shown for clarity (see text). }
\label{lambda3}
\end{center}
\end{figure}

The wavelength position of K3, $\lambda_{K3}$, and the corresponding  
core intensity $I_{K3}$ can be determined using a quadratic fit
over points centered around the pixel with minimum intensity.
The choice of the number of pixels to be used in the fit plays a critical role in
determining the quality of the result. The high spectral resolution of the ISS
requires a relatively large number of pixels 
to properly characterize the core of K3, which is quite
flat. For the new version of the reduction code we have used a total of 19 pixels. The result
is a more stable determination of $\lambda_{K3}$ as shown in the bottom left panel of
Figure \ref{lambda3}. On the other hand, the impact on $I_{K3}$ is not significant.

\subsection{1 \AA~ and 0.5 \AA~ emission index}

\begin{figure}
\begin{center}
\includegraphics[width=1.0\textwidth]{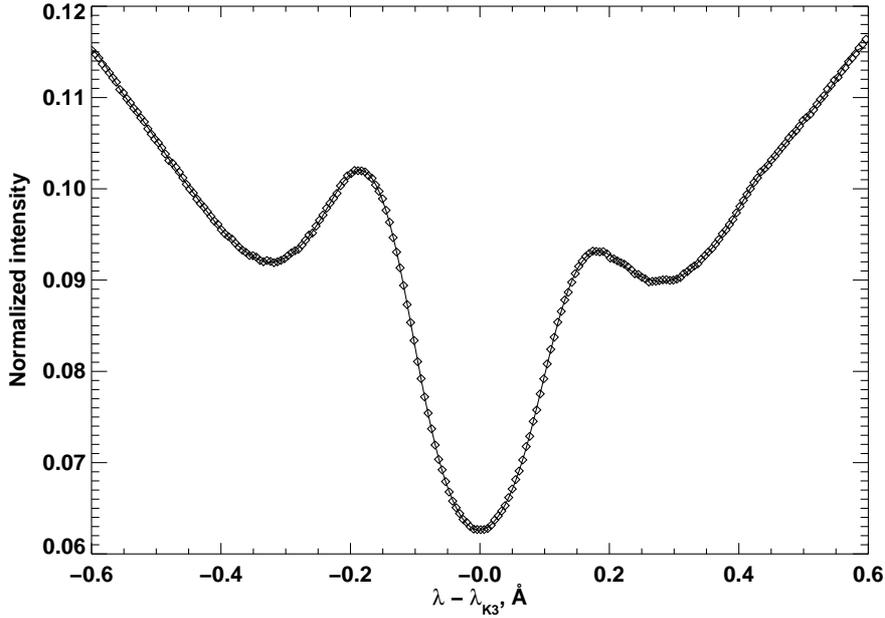}
\caption{Typical example of an observed ISS CaK profile (diamonds) reconstructed using 15
Fourier components (solid line). The wavelength scale is measured with respect to the center
of K3.}
\label{prof}
\end{center}
\end{figure}

To calculate the 1 \AA~ and 0.5 \AA~ emission index we have used a new approach based on
the Fourier components of the line profile. The central part of the \caii profile, between
-0.6\AA~ to +0.6\AA~from the center of K3, can be very well reconstructed using a limited
number of Fourier components:
$$I(x) = A_0 + 2\sum_{i=1}^{NC}\left[A_i\cos(2\pi i x/N) + B_i\sin(2\pi i x/N)\right],$$
where $x = -N/2 + j ~~(j = 0, N - 1)$, $N$ is the number of spectral samples, and $NC$ = 15 is
the number of components used in reconstructing the profile. The coefficients $A_i$ or $B_i$
can be easily determined by multiplying both sides of the above equation by $\cos(2\pi k x/N)$
or $\sin(2\pi i x/N)$ and summing over $x$, then making use of the property that sines and
cosines are orthogonal. An example of an observed and reconstructed profile is shown in Figure \ref{prof}.

\begin{figure}
\begin{center}
\includegraphics[width=1.0\textwidth]{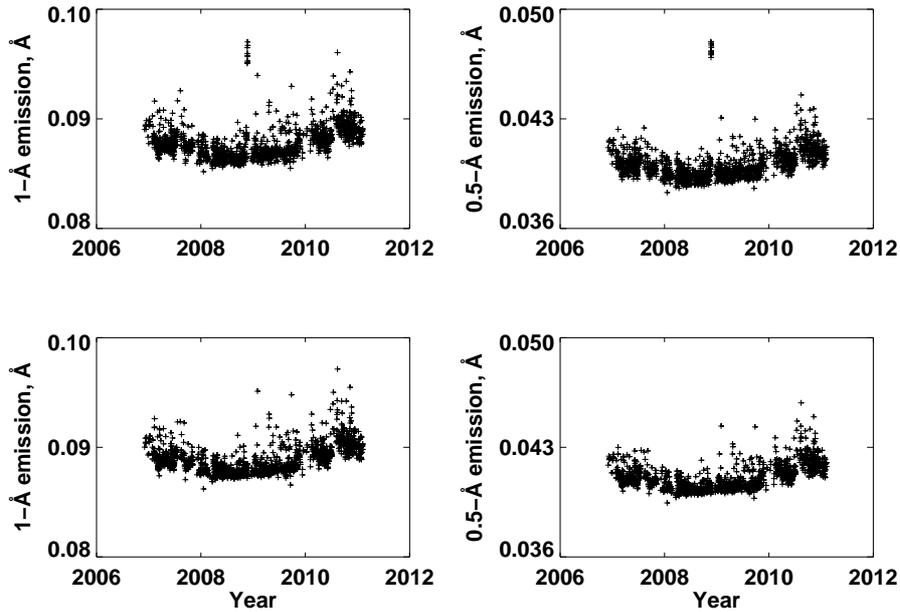}
\caption{
Comparison between CaK emission indices time series computed using a standard numerical integration scheme
(trapezoid rule, top two plots) and the analytical formula described in this document (bottom two plots).
The latest provides slightly higher values for the two emission indices. The error bars are not shown
for clarity (see text).
}
\label{em}
\end{center}
\end{figure}

The major advantage of using this representation is that the above equation can be integrated
analytically to compute the area under the line profile without the limitations in accuracy 
imposed by standard numerical integration schemes.
Therefore, the Ca K $EM$ emission parameter is given by:

$$
   EM = A_0(x_2 - x_1)+2\sum_{i=1}^{NC} \{(A_i/i)\sin[\pi i (x_2-x_1)/N]\cos[\pi i (x_2+x_1)/N]\} +
$$
$$
    -2\sum_{i=1}^{NC}\{(B_i/i)\sin[\pi i (x_2+x_1)/N]\cos[\pi i (x_2-x_1)/N]\},
$$

where $x_1$ and $x_2$ are the integration limits. For example, in case of
the 1 \AA~ emission, $x_1$ and $x_2$ are
the values corresponding to -0.5 \AA~ and +0.5 \AA, respectively.
The resulting time series are shown in Figure \ref{em}. 
The new reduction indicates
a slight increase in both the computed parameters.
This is because the typical trapezoid rule applied for integration underestimates the actual
area under the profile. 

\subsection{K2 and K1 differences}

\begin{figure}
\begin{center}
\includegraphics[width=1.0\textwidth]{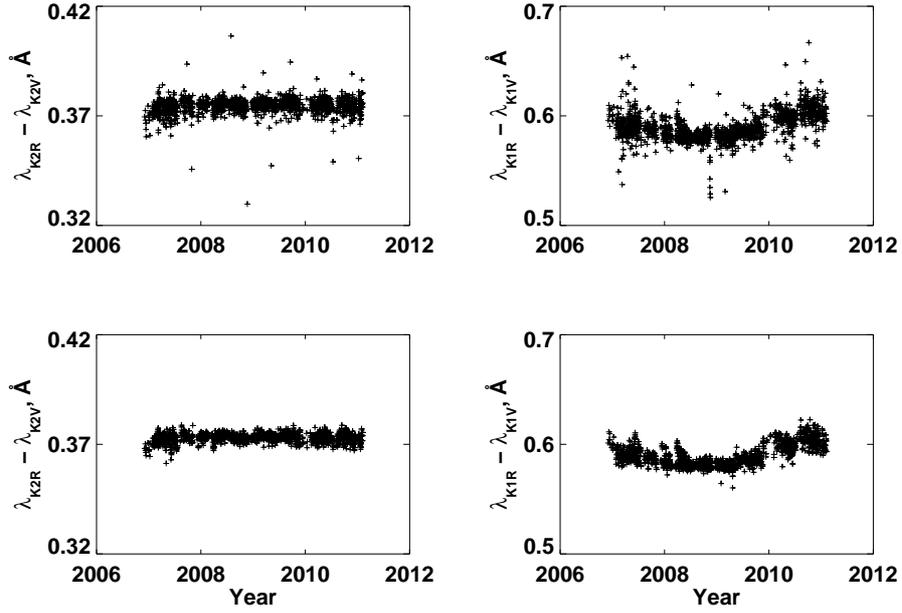}
\caption{Time series of wavelength separation of R and V emission K2 (left)
and K1 (right). Shown is a comparison between the results obtained from the
old ISS pipeline (top row) and the new calculation described in this article (bottom
row). The error bars are not shown for clarity (see text).}
\label{kvrdif}
\end{center}
\end{figure}

The wavelength positions of K1R ($\lambda_{K1R}$), K1V ($\lambda_{K1V}$),
K2R ($\lambda_{K2R}$), and K2V ($\lambda_{K2V}$) can be calculated using an approach
similar to the one used to compute $\lambda_{K3}$. However, in order to improve
the quality of the result we included an additional step just before performing the quadratic
fit. We take a 21-point interval centered around the minimum/maximum of K1/K2, and
we fit this portion of the spectrum with a 6th order polynomial. Instead of using
points from the observed spectrum, we use points
from this polynomial for our 7-point quadratic fit around the positions of K1R, K1V,
K2R, and K2V. The result is a more stable determination of the four wavelengths, as
clearly shown in Figure \ref{kvrdif}.

Once these four wavelengths have been determined, the calculation of the four corresponding
intensities is performed as described in subsection 4.2.

\subsection{K2-K3 asymmetry and K2V/K3 ratio}

These two parameters are derived from K3, K2R and K2V intensities:

$$\mbox{K2V/K3 } = I_{K2V}/I_{K3}$$

$$\mbox{K2-K3 asymmetry } = (I_{K2V} - I_{K3})/(I_{K2R} - I_{K3})$$

\begin{figure}
\begin{center}
\includegraphics[width=1.0\textwidth]{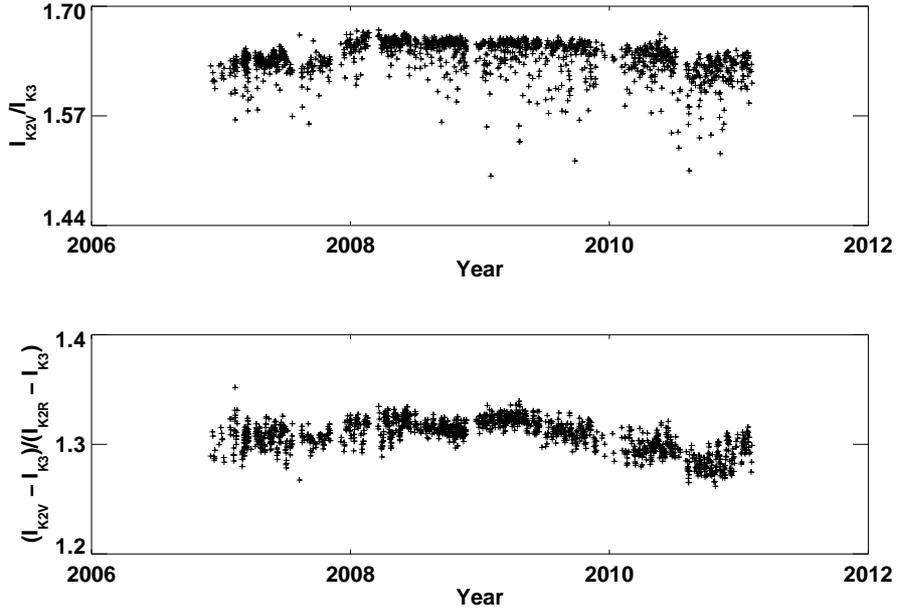}
\caption{Ratio of K2V to K3 intensities and line asymmetry ratio of V and R
parameter time series derived from the new ISS pipeline. The error bars are not shown for clarity (see text).}
\label{asym}
\end{center}
\end{figure}

These two parameter series marginally benefit from the new and improved version
of the code, as shown in Figure \ref{asym}.

\subsection{Wilson - Bappu index}

\begin{figure}
\begin{center}
\includegraphics[width=1.0\textwidth]{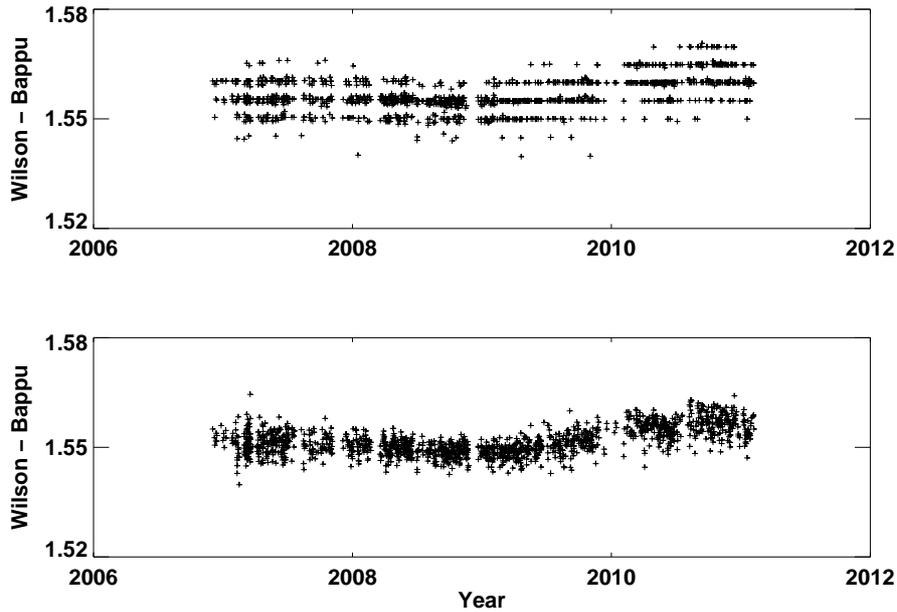}
\caption{The calculated Wilson - Bappu index time series using the old ISS
pipeline (top) and the new improved version (bottom).
Errors are not computed for this parameter.}
\label{wb}
\end{center}
\end{figure}

This index, $WB$, is defined as:

$$WB = \log[76.28(\lambda_{R} - \lambda_{V})],$$

where $\lambda_{R}$ and $\lambda_{V}$ are the wavelength positions corresponding
to $(I_{K1R}+I_{K2R})/2$ and $(I_{K1V}+I_{K2V})/2$, respectively. In the original
code these two positions were determined only to the rounded pixel. In this new version
we use a proper interpolation to accurately estimate these two wavelengths. A comparison
between these two versions of the code is shown in Figure \ref{wb}. Note that the
pixel-rounding problem clearly visible in the old time series has now been eliminated.

\section{Error estimate}

\begin{figure}
\begin{center}
\includegraphics[width=1.0\textwidth]{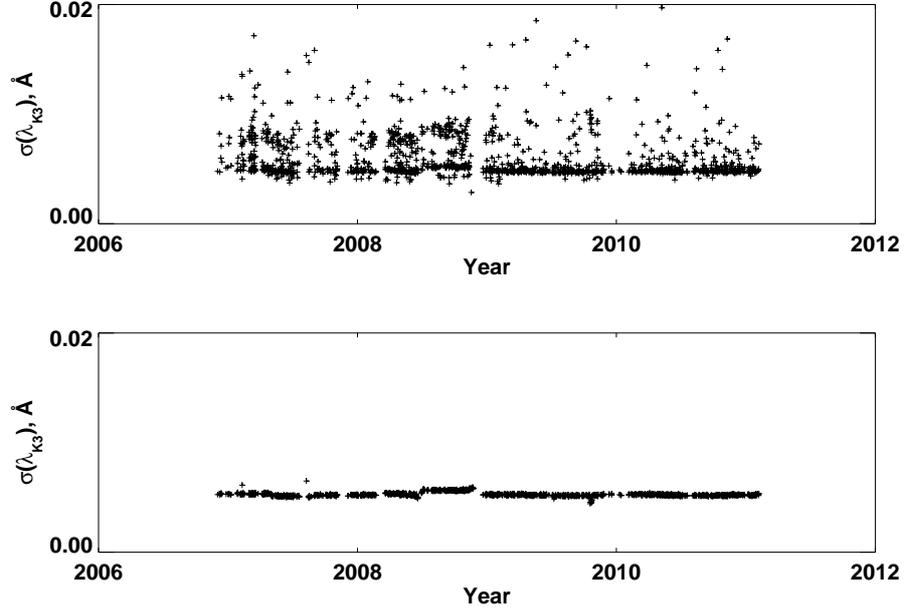}
\caption{Comparison between the formal 1-$\sigma$ error time series for $\lambda_{K3}$
calculated using the old ISS
pipeline (top) and the new improved version (bottom).}
\label{err}
\end{center}
\end{figure}

The original formulas used to calculate the errors associated with each parameter have not been
changed in this new version of the pipeline. However, the error determination has greatly
benefited from the improved estimation of the parameters. Figure \ref{err} illustrates a typical
comparison between the formal 1-$\sigma$ error time series computed for $\lambda_{K3}$ 
using the old ISS pipeline and the new improved calculation.  
The most important feature in this comparison is the much reduced scatter in the error
determination from one observation to another. The much more homogeneous behavior in the
error time series is achieved also for the other eight parameters. This result indicates that
the error for all parameters has been been reduced significantly for many observed spectra as
compared to the previous data reduction. This is particularly true for the measured wavelengths.
Finally, we should mention that errors are not reported for the Wilson - Bappu index.

\section{Conclusions}

Time series plots shown in the previous section indicate significant improvements in parameters derived from \caii line profiles. For eight parameters, error bars are much smaller for many data points as compared with previous data reduction (no error bars are computed for Wilson-Bappu parameter). 
Most importantly, we see very little variation in the amplitude of error bars throughout the five year observing period. Together with nearly constant linear dispersion, 
this lack of variation in the amplitude of errors indicates the extreme stability of the Integrated Sunlight Spectrometer.
The novel method of computing the emission indices based on the Fourier decomposition of 
the spectral line profile allows a more precise determination of these parameters. 
Comparison with widely adopted numerical integration schemes shows that these numerical 
integration methods may underestimate the emission measurements.

\begin{table}
\caption{Cross-correlation matrix generated from the nine parameter time series.
The indices correspond to: 1=Line intensity (K3), 2=1-Angstrom emission index,
3=Half-Angstrom emission index, 4=Ratio of K2V to K3 intensities, 5= Wavelength separation of V \& R emission maximum,
6=Wavelength separation of V \& R emission minimum, 7= Line asymmetry, 8=Wavelength for K3, and
9=Wilson-Bappu parameter.}
\begin{tabular}{rrrrrrrrrr}
\hline
 & 1 & 2 & 3 & 4 & 5 & 6 & 7 & 8 & 9 \\
\hline
1 &  1.000 &  0.992 &  0.983 & -0.931 & -0.337 &  0.655 & -0.548 & -0.629 &  0.520 \\
2 &  0.992 &  1.000 &  0.992 & -0.888 & -0.324 &  0.693 & -0.582 & -0.654 &  0.568 \\
3 &  0.983 &  0.990 &  1.000 & -0.854 & -0.379 &  0.761 & -0.638 & -0.709 &  0.611 \\
4 & -0.931 & -0.888 & -0.854 &  1.000 &  0.249 & -0.397 &  0.368 &  0.433 & -0.295 \\
5 & -0.338 & -0.324 & -0.379 &  0.250 &  1.000 & -0.362 &  0.214 &  0.362 & -0.161 \\
6 &  0.655 &  0.693 &  0.761 & -0.397 & -0.362 &  1.000 & -0.820 & -0.809 &  0.768 \\
7 & -0.548 & -0.582 & -0.638 &  0.368 &  0.214 & -0.820 &  1.000 &  0.795 & -0.694 \\
8 & -0.629 & -0.654 & -0.709 &  0.433 &  0.362 & -0.809 &  0.795 &  1.000 & -0.647 \\
9 &  0.520 &  0.568 &  0.611 & -0.295 & -0.161 &  0.768 & -0.694 & -0.647 &  1.000 \\
\hline
\end{tabular}

\label{iss_ccm}
\end{table}

Routine observations with the ISS were started at the end of 2006 at the declining phase of Solar Cycle 23. 
Although the scientific study of this data set is outside the scope of
the current article, one can note clear
cycle variations in several parameters of this spectral line with a
minimum months earlier than the official
sunspot minimum of December 2008. 
The K3 line core intensity, the two $EM$ emission parameters, and the 
wavelength separation of V \& R emission minimum
show a minimum around August-September 2008 ($\pm$ 40 days).
The Wilson - Bappu index time series has an even earlier minimum, around the end of June 2008
($\pm$ 40 days).
Such mismatchs
between solar cycle minima in different indices has been reported in
earlier papers (e.g., \opencite{2002AnGeo..20..741K}). However, \inlinecite{2002AnGeo..20..741K} had found that
Ca K-line intensity is in good agreement with sunspot number in timing of
solar minimum, while our data show Ca II K3 minimum intensity preceding sunspot
minimum. 

Finally, to better understand the relationship between pairs of parameters we
have generated the cross-correlation matrix from the nine time series.
This cross-correlation matrix is shown in Table \ref{iss_ccm}.
It appears that K3 intensity, 0.5-\AA~ and 1-\AA~ emission indices are highly correlated with each other.
Wavelength separation of V \& R emission maximum does not correlate well with any other parameters. K3 wavelength shows moderate correlation with most other parameters, and in fact, it varies inversely with solar cycle.

\begin{acks}
The NSO and NOAO are operated by the Association of University for Research in Astronomy,
Inc. (AURA), under cooperative agreements with the National Science Foundation.
\end{acks}



\end{article} 
\end{document}